\documentclass[aps,twocolumn,superscriptaddress,floatfix,nofootinbib,prl]{revtex4-1}

\usepackage{lipsum}
\usepackage{graphicx}
\usepackage{amsmath,amssymb}
\usepackage{braket}
\usepackage{mathtools}
\usepackage{hyperref}
\usepackage{multirow}
\usepackage{color}
\usepackage[normalem]{ulem}
\usepackage{amsfonts}
\usepackage{float}
\usepackage{pdfpages} 
\makeatletter

\AtBeginDocument{\let\LS@rot\@undefined}
\makeatother

\begin{document}
\newcommand{\lnJ}[2]{\ensuremath{\ell^\text{J}_{{#1}{#2}}}}
\newcommand{\lnh}[1]{\ensuremath{\ell^\text{h}_{{#1}}}}
\newcommand{\V}[1]{\vec{{#1}}}
\newcommand{\RGerror}{\Delta_\text{RG}}
\newcommand{\Mag}{\mu_\text{M}}
\newcommand{\parameter}{g}
\newcommand{\flag}[1]{\textcolor{red}{#1}}

\title{Quantum criticality in the 2d quasiperiodic Potts model}
\author{Utkarsh Agrawal}
\affiliation{Department of Physics, University of Massachusetts, Amherst, MA 01003, USA}
\author{Sarang Gopalakrishnan}
\affiliation{Department of Physics and Astronomy, CUNY College of Staten Island, Staten Island, NY 10314;  Physics Program and Initiative for the Theoretical Sciences, The Graduate Center, CUNY, New York, NY 10016, USA}
\author{Romain Vasseur}
\affiliation{Department of Physics, University of Massachusetts, Amherst, MA 01003, USA}
\begin{abstract}

Quantum critical points in quasiperiodic magnets can realize new universality classes, with critical properties distinct from those of clean or disordered systems. Here, we study quantum phase transitions separating ferromagnetic and paramagnetic phases in the quasiperiodic $q$-state Potts model in $2+1d$. Using a controlled real-space renormalization group approach, we find that the critical behavior is largely independent of $q$, and is controlled by an infinite-quasiperiodicity fixed point. The correlation length exponent is found to be $\nu=1$, saturating a modified version of the Harris-Luck criterion. 

\end{abstract}
\vspace{1cm}

\maketitle

Quenched disorder can dramatically affect the universality class of a quantum phase transition, and drive it to a new renormalization group (RG) fixed point if the correlation length exponent $\nu$ violates the Harris criterion $\nu \geq 2/d$~\cite{Harris1974,Chayes1986a} with $d$ the dimensionality of the system. As the effective randomness grows under renormalization, the new infrared fixed point can either be characterized by finite or infinite randomness. 
Infinite-randomness fixed points can be analyzed using an asymptotically exact real space renormalization group (RSRG) approach~\cite{Ma1979, Fisher1992,XXZ_Fisher_RG} that yields exact predictions for critical exponents and scaling functions. The RSRG approach has been applied to many different quantum phase transitions in one and two dimensions, both at zero temperature and in the context of many-body localization~\cite{Fisher1992,Ma1979,fisher1995critical, fisher1999phase,XXZ_Fisher_RG, motrunich_dynamics_2000,motrunich_infinite-randomness_2000,Senthil1996,hyman1997, PhysRevLett.93.150402, refael_moore, igloi2005strong,kovacs_critical_2009,kovacs_renormalization_2010,PhysRevLett.89.277203,Laumann2012,lin_entanglement_2007,Fidkowski2008,Bonesteel2007,zzdh, gvs, dgpsv, mh2019, mhi, vha, pvp, dvp, thmd, PhysRevX.4.011052, PhysRevLett.110.067204, QCGPRL, PhysRevB.93.104205, GOPALAKRISHNAN20201}. 

The structure of infinite-randomness critical points depends crucially on the assumption of spatially uncorrelated disorder. However, 
many present-day experiments, involving, e.g., twisted bilayer graphene~\cite{GrapheneBistritzer2011,GrapheneCao2018a,GrapheneCao2018} and ultracold atoms in bichromatic laser potentials~\cite{ColdBordia2017,ColdDeissler2010,ColdLuschen2017,ColdRoati2008,ColdSchreiber2015} 
involve systems that are spatially inhomogeneous, but quasiperiodic rather than random. Quasiperiodic potentials are deterministic, with strong spatial correlations, so they do not lead to conventional infinite-randomness behavior~\cite{Yoo2020,Yao2020,Baboux2017,Verbin2015,Lahini2009,Tanese2014,Deguchi2012}.
Instead, when a clean critical point is unstable to quasiperiodicity, it flows to a new class of fixed points. Field theoretic methods~\cite{Wiseman1998,Aharony1996,Fischer1977,Boyanovsky1982,Kim1994,Ye1993,XXZ_fisher_perturbative} do not easily generalize to quasiperiodic systems~\cite{vidal_correlated_1999,vidal_interacting_2001}, because there is no disorder to average over. However, very recent results~\cite{Crowley2018a,Crowley2018,Agrawal2019} have revealed the existence of ``infinite-quasiperiodicity'' quantum critical points~\cite{Agrawal2019} in one dimensional spin chains; at these critical points, RSRG yields exact predictions for exponents. 
Despite their differences, infinite-quasiperiodicity and infinite-randomness critical points share the key feature that the dynamical critical scaling exponent $z = \infty$: thus, the characteristic timescale $t_\xi$ associated with a length-scale $\xi$ grows faster than any power law of $\xi$.
So far, such infinite-quasiperiodicity fixed points have chiefly been studied in one dimension; higher-dimensional cases are poorly understood~\cite{Sbroscia2020,Inoue2020,Szabo2020,QuantumSpins_Jagannathan}. The $z = \infty$ dynamical scaling leads to a rapidly vanishing gap, which makes it hard to access the critical regime using Quantum Monte Carlo techniques~\cite{Pich1998,Guo1994,Rieger1994a,2020arXiv200809617K}. Tensor network based approaches (see {\it e.g.}~\cite{ORUS2014117}) are also less suited to study 2d QP quantum criticality, due to large entanglement. 

In this letter, we propose a general RSRG approach to study 2+1d quantum spin models with QP couplings. As in the implementations of RSRG for disordered systems in two dimensions,  the RG changes the underlying geometry of the system creating intricate and complex long range interactions~\cite{motrunich_infinite-randomness_2000,kovacs_renormalization_2010,lin_entanglement_2007}. Nevertheless the RG procedure can be efficiently implemented numerically. We focus on the 2d quantum Potts model, with $q$ ``colors'' ($q=2$ corresponding to the Ising model). For clean systems, the phase transition separating paramagnetic and symmetry-broken phases is in the classical 3D Potts model universality class, which is a first-order for $q\geq 3$~\cite{Bazavov2008,Hellmund,Wu1982}. Strong enough QP modulations should smoothe these first-order transitions~\cite{Hui1989}, driving them to a new strong quasiperiodicity fixed point that we describe using RSRG. Our results suggest that the critical properties do not depend on $q>2$, with the Ising case $q=2$ being special. 
Beyond our numerical results for the critical exponents, we 
propose a general argument for the correlation exponent $\nu=1$ for these new infinite-quasiperiodicity transitions, based on the distribution of ``defects'' in the critical structure. Due to the deterministic and almost periodic nature of quasiperiodic potentials these defects form a definite pattern; in some special cases, the defects form a QP tiling with a length scale that defines the correlation length.
Interestingly, the value of $\nu$ saturates a modified version of the Harris-Luck criterion~\cite{Luck1993}, namely $\nu \geq 1$; the modifications are due to boundary fluctuations coming from correlations in boundaries of rectangular patches at \emph{all} length scales. 

\paragraph{{\bf Model.}} The $q$-state quantum Potts model is defined via the Hamiltonian
\begin{equation}
H=-\sum_{\langle i,j\rangle} J_{ij}\delta_{n_i,n_j} - \sum_i \frac{h_i}{q}\sum_{n_i,n_i'}|n_i\rangle \langle n_i'|, \label{eq: potts hamiltonian}
\end{equation}
defined on the square lattice with $\langle i,j\rangle$ denoting nearest neighbor pairs, where $n_i$ is a variable on site $i$ that takes one of $q$ possible values. The first term with $J_{ij}>0$ is a classical ferromagnetic interaction favoring aligned spins, while the second term is a quantum transverse field leading to a paramagnetic phase at large $h_i$'s. For $q=2$ colors, this coincides with the familiar transverse field Ising model. The model is initially defined on a square lattice; however, we believe our results to be independent of the initial lattice geometry, as RSRG drastically changes the connectivity of the system.

The couplings $J_{ij}>0, h_i>0$ are inhomogeneous, aperiodic but deterministic. Here, we consider $J_{ij} = f_1(\V{k}_1 . \V{r})+f_2(\V{k}_2.\V{r})$, where $\V{r}=(i_x,i_y)+\frac{1}{2}(j_x-i_x,j_y-i_y)$, $\V{k}_1$, and $\V{k}_2$ are two orthogonal unit vectors, and $f_a(x)=f_a(x+\varphi^{-1})$ for some irrational $\varphi$, which we take to be the golden ratio, $\varphi = \frac{1+\sqrt{5}}{2}$. Similarly, the fields are taken from an initial potential of the form, $h_i = g_1(i_x)+g_2(i_y)$ with $g_a(x)=g_a(x+\varphi^{-1})$. For concreteness, we focus on the following QP modulations throughout the paper,
\begin{align}
\lnJ{i}{j}=&2+\cos{(2\pi\varphi\V{k}_1. \V{r}+\phi_1)} + \cos{(2\pi\varphi\V{k}_2 . \V{r} + \phi_2)} \label{eq: potential} \\ \nonumber
\lnh{i} =& \parameter (2+\cos{(2\pi\varphi i_x + \phi_3)} + \cos{(2\pi\varphi i_y + \phi_4)}), 
\end{align}
where $g$ is a parameter driving the transition, $\lnJ{i}{j}=-\ln J_{ij}$ and $\lnh{i}=-\ln h_{i}$ are defined so as to decrease the transient behavior in the RG (see below), and $\phi_i$ are some constant global phases which we average over. Unless otherwise stated, we take $\vec{k}_1=(\sin\theta,\cos\theta)$, with the angle $\theta=\sqrt{2}\pi$. Our results do not depend on the details of these distributions~\cite{supmat}.

\paragraph{\bf RG procedure.}  We now describe the RSRG procedure we use to capture the critical properties of Eq.~\eqref{eq: potts hamiltonian}. One step of the RG procedure consists of identifying the strongest coupling in the Hamiltonian (which sets the cutoff, $\Omega$) and eliminating it, as follows \cite{motrunich_infinite-randomness_2000,kovacs_critical_2009,kovacs_renormalization_2010,Yu2008}. If the strongest coupling is a bond $J_{ij}$, one merges the two spins connected by the bond into a new effective spin (or ``cluster'') with magnetic moment $\mu_i' = \mu_i + \mu_j$ ($\mu_i =1$ for initial physical spins). The effective transverse field acting on the cluster is given by second-order perturbation theory, $h_i'\approx\frac{h_i h_j}{\kappa J_{ij}}$ with $\kappa=q/2$; also, any other spin (or cluster) in the system that was connected to either $i$ or $j$ now picks up a bond to the new cluster, with coupling given by $J_{ik}'=\text{max}(J_{ik},J_{jk})$. If instead the strongest spin is an effective field $h_i$, one eliminates the site $i$. Any other pair of sites $j, k$ that were connected to $i$ by bonds now pick up a new effective bond, which we estimate using 2nd order perturbation theory: $J_{jk}' \approx J_{jk}+\frac{J_{ij}J_{ik}}{\kappa h_i}\approx\text{max}(J_{jk},\frac{J_{ij}J_{ik}}{\kappa h_i})$.
This procedure correctly captures the low energy physics as long as $\Omega \gg J_{ij},h_j$ (broadly distributed couplings) so that perturbation theory is controlled; we will see that for infinite-quasiperiodicity fixed points, the parameter controlling the error in perturbation theory flows to zero upon coarse-graining, leading to asymptotically exact predictions for universal properties.

\begin{figure}[t]
\centering
\includegraphics[scale=0.42]{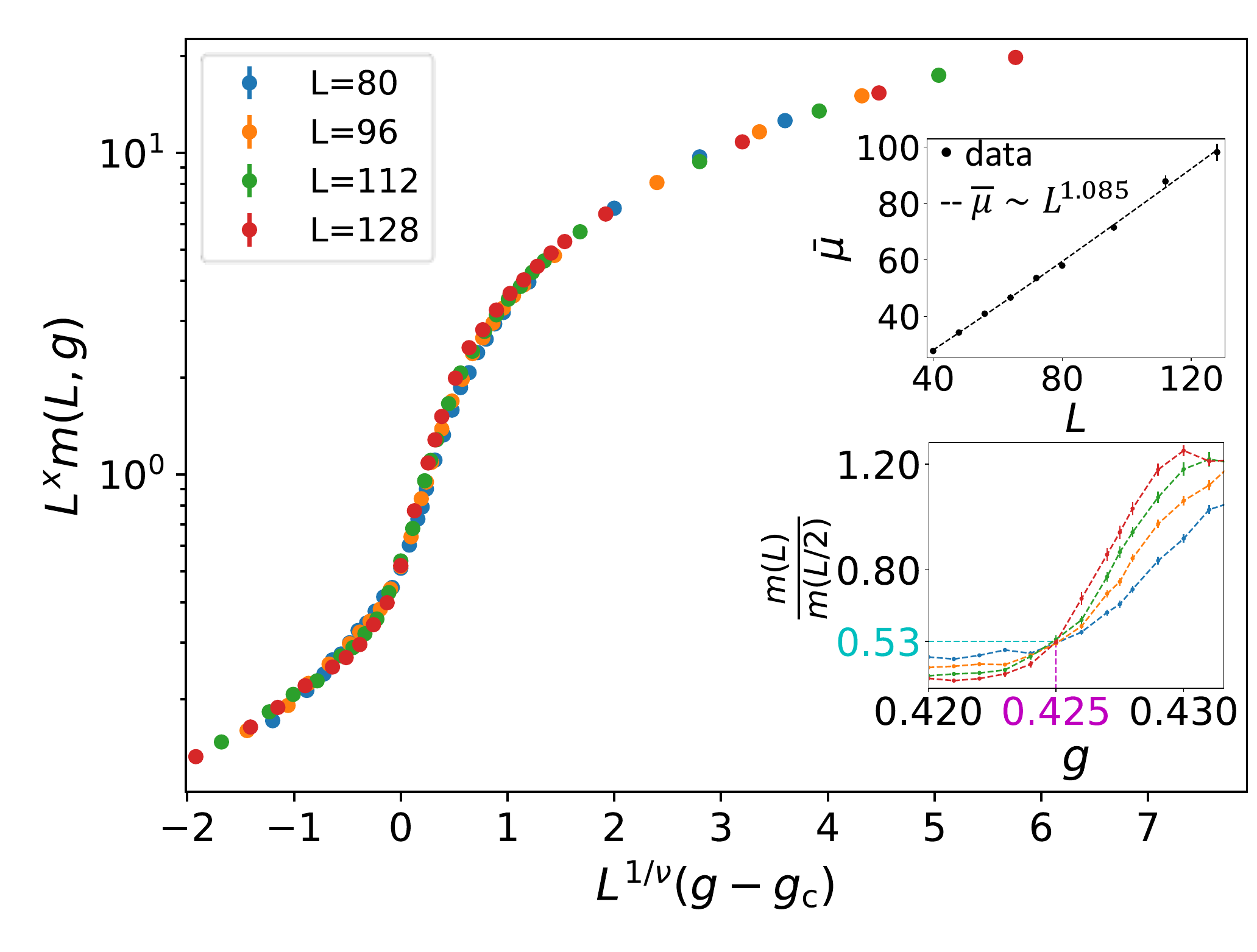}
\caption{{\bf Magnetization scaling.} Scaling collapse of the magnetization $m(L,g)$ for $q=3$ with the correlation length exponent $\nu=1$, critical coupling $g_c=0.425$, and magnetization scaling dimension $x=0.92$. \emph{Bottom inset:} Plot of the ratio $r(L)=\frac{m(L)}{m(L/2)}$ vs $g$. In the para- and ferromagnetic phases $r(L)$ depends on $L$ (large $g$ corresponds to a ferromagnet, small $g$ to a paramagnet), while at the critical point this ratio is a constant. Defining the scaling dimension $x$ via $m\sim L^{-x}$, we have $2^{-x}\approx 0.53$ or $x\approx 0.92$. The critical point is $g_c = 0.425$. \emph{Top inset:} Average magnetic moment $\overline{\Mag}$ vs $L$ giving $\overline{\Mag}\sim L^{d_f}$ with $d_f=1.085\pm  0.024$. This is consistent with $x+d_f = 2$.}\label{Fig: ratio and collapse}
\end{figure}

We numerically run the RG procedure described above starting from a $L\times L$ square lattice.  We first focus on the $q=3$ Potts model -- the critical behavior is largely independent of $q \geq 3$. As the system moves along the RG flow, its geometry changes giving rise to graphs of increasingly intricate connectivity. Instead of implementing the RG in the naive sequence described above (i.e., always decimating a single largest coupling), we follow standard techniques~\cite{kovacs_renormalization_2010} to optimize the decimation sequence. (We have checked that at the end of the RG procedure, the optimized and naive decimation sequences yield identical couplings, so this step is \emph{not} an approximation.)

\begin{figure}
\centering
\includegraphics[scale=0.45]{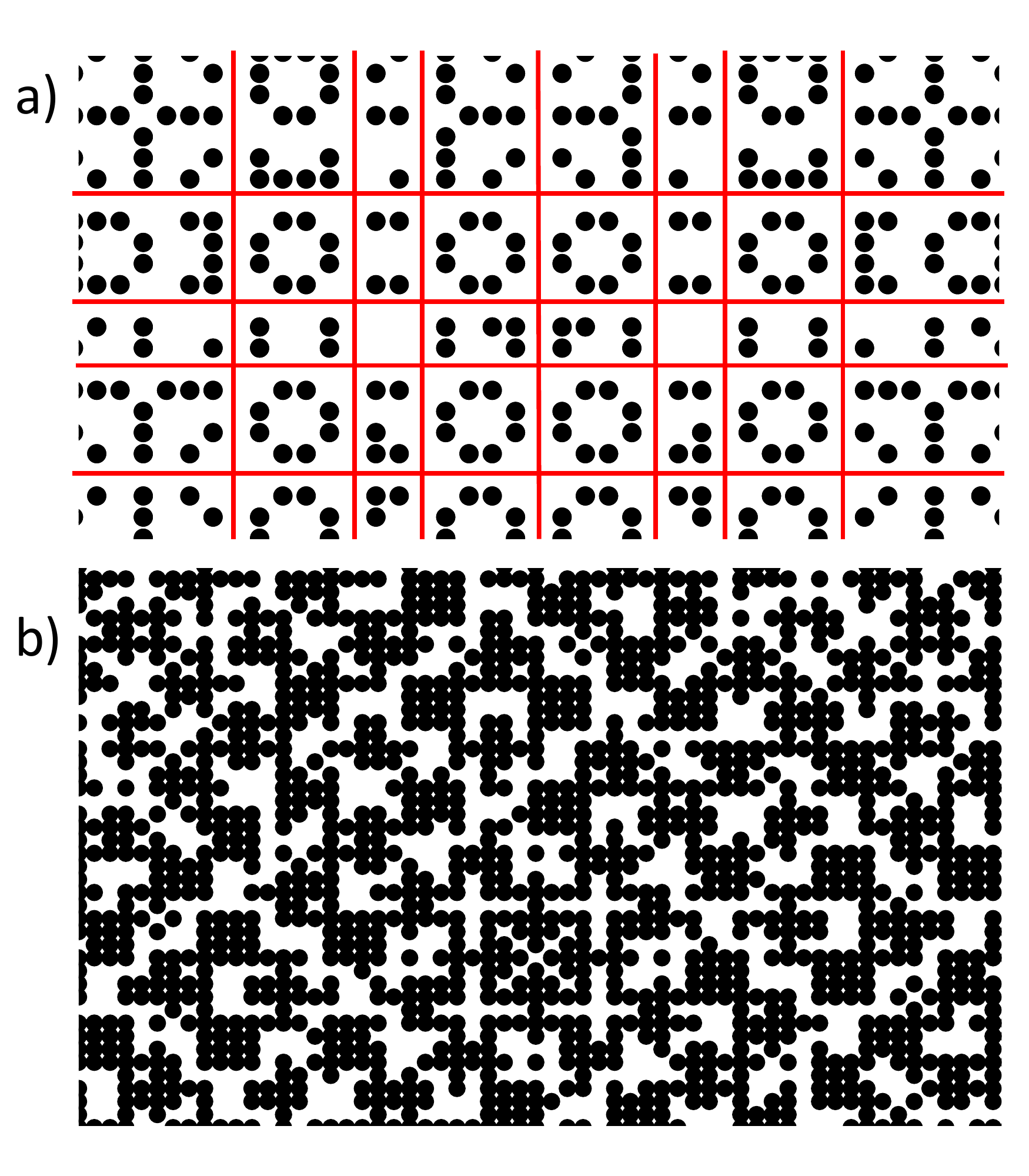}
\caption{{\bf Critical defects and quasiperiodic tiling structure.} \emph{a}) Geometry of the set $S = \{ i: \ \min\{\lnJ{i}{j}\} < \lnh{i} \}$ where $\lnh{}$, $\lnJ{}{}$ are defined in \eqref{eq: potential} with the angle $\theta=0$. We have taken $g=0.4$ for illustration purposes. Black sites belong to $S$, while white sites do not, and form single-site clusters. We see pockets of black sites separated by 1d section of white sites, marked by red lines. These red lines form a square QP tilling. Large clusters in later steps of the RG are formed by joining small clusters within different tiles/faces of the red lattice. Defects are breaks in the pattern of inter tile connections away from the critical point. The number of breaks are proportional to the inverse of detuning parameter $\delta$, giving $\nu=1$.  \emph{b}) Geometry of $S$ for $g=g_c=0.425$ and $\theta=\sqrt{2}\pi$. The structure is not as clear and well defined as in the $\theta=0$ case but we still see local puddles in $S$. } \label{tilling}
\end{figure}

\begin{figure*}[t]
\begin{center}
\includegraphics[width=\textwidth]{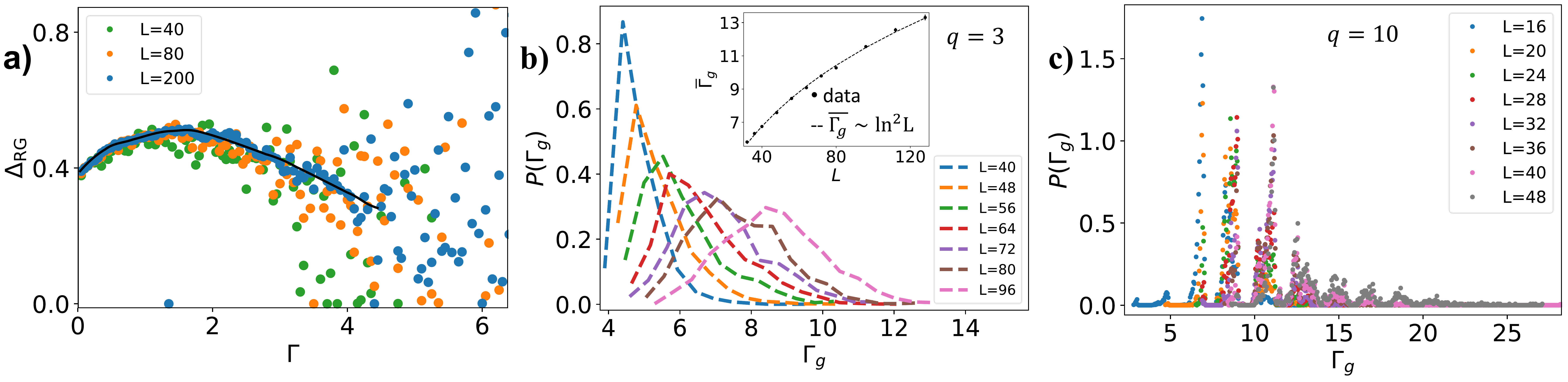}
\caption{{\bf RG errors and gap distribution.} \emph{a}) Plot of the RG error, $\RGerror$, vs RG time $\Gamma(\equiv -\ln\Omega)$ at the critical point. Data from 9 different phase realizations are combined and averaged over windows of $\Gamma$ of size 0.05. We see a trend of the error decreasing with the RG, i.e increasing $\Gamma$ (the black curve is a guide for the eye), whereas towards the end of the RG the data becomes more scattered and noisy. As we increase system sizes, the onset of the data scattering shifts towards latter stages of the RG, consistent with the noisiness in the error at higher $\Gamma$ being a finite size effect. \emph{b}) Distribution of logarithmic of gap for $q=3$, $-\ln \Delta E_g\equiv \Gamma_g$.  With increasing system size, the average is increasing with the distribution becoming broader, indicating a broadening of couplings and fields along the RG flow. {\it Inset:} Scaling of the finite-size gap, showing $\overline{\Gamma}_g$ vs $L$; the fit is compatible with $\overline{\Gamma}_g\sim \ln^2 L$. Binning window for $\Gamma_g$ was taken to be $0.5$. \emph{c}) Distribution of logarithmic of gap for $q=10$ with window size of $0.05$. Unlike the $q=3$ case, we see a systematic rise and fall in $P(\Gamma_g)$, with the probability going to zero for some values of the gap. This is reminiscent of the 1d case where a similar banding of couplings and gaps was observed~\cite{Agrawal2019}.} \label{Fig. RG error}
\end{center}
\end{figure*}

\paragraph{{\bf Magnetization and fractal exponent.}}
At the end of the RG, the surviving cluster with moment $\Mag$ determines the magnetization of the system, $m(L,g)=\Mag /L^2$, where $L$ is the linear size of the system. To locate the critical point we plot $r(L,g)=m(L,g)/m(L/2,g)$ vs $g$ for various $L$; away from the criticality $r(L,g)$ changes with L, while being scale independent at the critical point~\cite{Yu2008}. The critical magnetization scales as $m(L,g_c)\sim L^{-x}$ giving the crossing value $r(L,g_c)=2^{-x}$. The average moment of the cluster at the critical point scales as $\Mag \sim L^{d_f}$ with $d_f$ being the fractal dimension of the spins in the cluster. Those two exponents satisfy the scaling relation $d_f + x = 2$.
Those quantities are plotted for the $q=3$ Potts model in Fig.~\ref{Fig: ratio and collapse}, and we find $2^{-x}\approx 0.53$ or $x\approx 0.92$ and $d_f = 1.085 \pm 0.024$, consistent with the relation $d_f + x = 2$.

\paragraph{\bf Correlation length.} Assuming single parameter scaling with a diverging correlation length $\xi \sim |g - g_c|^{-\nu}$, we expect the following scaling form for the magnetization $m(L,g)=L^{-x}f((g-g_c)L^{1/\nu})$, where $f$ is a universal scaling function. Using the values of $g_c$, and $x$ obtained from the plot of $r(L)$, we find a nice collapse for $\nu \approx 1$. We now argue that this result $\nu = 1$ holds exactly, at least for some classes of quasiperiodic potentials. 

The argument for $\nu = 1$ is as follows. Let us first consider the case where the quasiperiodic modulation is parallel to the lattice vectors, i.e., $\V{k}_1=(1,0)$, $\V{k}_2=(0,1)$ in~\eqref{eq: potential}. We now consider running the RG for two realizations of the lattice, one at criticality and one detuned by a distance $\delta$. We now look for ``defects,'' or points on the lattice where the two RG realizations begin to diverge (because one of them decimates fields and the other bonds). Defects occur when locally, fields are close ($\alt \delta$) in magnitude to the neighboring bonds; thus, a small detuning is enough to change the order of decimations. However, because the quasiperiodic structure is approximated to precision $\sim \delta$ by a rational approximant with period $\sim 1/\delta$, each defect has an almost perfect repeat at a distance $\sim 1/\delta$ (along both lattice directions). This can be seen by observing that $\cos{(2\pi\varphi (x+F_n)+\phi)} = \cos (2\pi\varphi x + \phi) + \mathcal{O}(\varphi^{-n})$, where $F_n$ is the nth Fibonacci number: defects must repeat along the vertical and horizontal axis, forming a QP tilling, with a length scale $\xi=F_n\sim\varphi^n$, with $\delta\sim\varphi^{-n}$ giving $\xi\propto \delta^{-1}$ (see~\cite{Agrawal2019} for a similar argument in quantum spin chains). 
Thus, when the RG reaches length scale $1/\delta$, defects will proliferate and drive the system away from criticality, corresponding to $\nu=1$. To illustrate this tiling geometry, we plot the set $S = \{ i: \ \min\{\lnJ{i}{j}\} < \lnh{i} \}$, where the $\min$ is over nearest neighbors. This condition is satisfied for couplings $J$ that are decimated first in the RG, forming non-trivial clusters. The geometry of the set $S$ is shown in Fig.~\ref{tilling}. 

The geometry away from $\theta = 0$ is less transparent, but numerics once again suggests $\nu = 1$; moreover, the model remains strongly anisotropic under coarse-graining, with preferred orientations (Fig.~\ref{tilling}b). We now argue that, if this anisotropy persists under the RG, it leads to a modification of the Harris-Luck bound on $\nu$~\cite{Luck1993}. The standard argument for this criterion runs as follows. In a large patch of the sample of linear dimension $\ell$, the apparent local value of the critical point is $\delta_\ell \equiv \langle g \rangle_\ell - g_c \sim \ell^{w-d}$ where $w$ is the wandering exponent. Setting $\ell$ to the correlation length $\xi \sim \delta^\nu$, we get $\delta_\xi \sim \delta^{\nu(d-w)}$. When $\delta_\xi$ is small compared with the global detuning $\delta$, the transition is well-defined. This criterion amounts to $\nu > 1/(d - w)$. Generic patches of a quasiperiodic system have wandering exponent $w = 0$ in the bulk so the standard Luck criterion reads $\nu > 1/d$. However, this analysis ignores ``boundary'' terms due to lines or other sub-dimensional regions of the sample where $\delta$ is locally away from its average value. If one includes these boundary contributions, the deviation is $\delta_\ell \sim \ell^{(d-1)-d} \sim 1/\ell$, so that $\nu \geq 1$ regardless of dimensionality. The quasiperiodic Potts model appears to saturate this modified bound,  with $\nu = 1$ (up to logarithmic corrections).

\paragraph{\bf Dynamical scaling and RG error.} We now turn briefly to the dynamical scaling properties at this transition. One can argue analytically that the timescale for a region of $\ell$ spins grows at least as $\ln t_\ell \agt \ln^2\ell$. This scaling follows naturally from the RG rules; recall that these rules involve a factor $\kappa > 1$ at each step. One can check that upon decimating a region of size $\ell$ to a single spin, one picks up at least $\ln^2 \ell$ factors of $\kappa$ in the effective couplings~\cite{supmat}, implying an energy scaling $- \ln E_\ell \sim \ln t_\ell  \gtrsim  \ln^2 \ell$. This scaling can be interpreted as the scaling of the finite size gap of a region of size $\ell$. 
 This divergence might be subleading (as it is in the random case), but guarantees ``activated'' scaling, where $t$ grows faster than any power of $\ell$. As we see in Fig.~\ref{Fig. RG error}b, our numerical results are consistent with $\ln t_\ell \sim \ln^2 \ell$, i.e., the same dynamical scaling as in one dimension~\cite{Agrawal2019}. We note that our data is also compatible with other types of activated scaling~\cite{supmat}.

A consequence of activated dynamical scaling is that the RG becomes increasingly accurate at late stages. The typical RG error (defined as $\log \RGerror \equiv \langle \log(\frac{\max J_{ij},h_i}{\Omega}) \rangle$, where the max function is over all neighboring terms of $\Omega$, with $\langle \cdot \rangle$ denoting average over a small window of $-\log \Omega$, and several phase realizations) vs $-\ln\Omega(\equiv \Gamma)$ at the critical point is plotted in Fig. \ref{Fig. RG error}. a). We see that on average, the RG error decreases along the RG flow, suggesting that the RG becomes asymptotically exact, as in the random case~\cite{motrunich_infinite-randomness_2000}. While the system sizes we can access remain away from the asymptotic regime where the RG is fully controlled, we observe very good quality critical data (Fig.~\ref{Fig: ratio and collapse}) with no signs of finite-size drifts.
Extrapolating these results, we expect the error of a typical RG step to go to zero asymptotically with $\Gamma$. 

\paragraph{\bf Critical behavior {\it vs} $q$.}

We conclude this letter by briefly discussing the case of $q>3$. 
For $q>3$, we observe a similar behavior as for $q=3$; there is a 2nd order transition with the RG becoming more controlled with the flow. The correlation exponent $\nu=1$ seems to hold, as expected from the general arguments discussed above. Unsurprisingly, the location of critical point is non-universal and changes with $q$ and $\theta$. The $d_f$ and $x$ exponents appear to be same for all values of $q>2$, suggesting the same universality class for different $q$'s, though we cannot exclude small differences based on our numerical data.  Interestingly, for larger values of $q$, we observe that the distributions of the gap and of couplings form ``bands'', with forbidden values in between the allowed bands (see Fig. \ref{Fig. RG error}.c). This is reminiscent of similar banding properties that were observed in QP quantum spin chains~\citep{Agrawal2019}; it would be interesting to investigate whether this can be leveraged to understand this RG analytically in the future.

The case of $q=2$ (the Ising model), is special. In this case, we find that the RG {\em does not} flow towards infinite quasiperiodicity, and is therefore not controlled. A similar scenario occurs in 1d \emph{weak} QP modulations are marginally irrelevant~\cite{PhysRevX.7.031061,Crowley2018a,Crowley2018,Agrawal2019} at the clean fixed points. However, unlike the 1d case, we observe that even on introducing strong QP modulations, the RG {\em does not} flow to infinite quasiperiodicity. From the modified version of the Luck criterion, we expect QP modulations to be relevant at the clean Ising transition, driving the system to a finite quasiperiodicity fixed point that cannot be described using RSRG. It would be especially interesting to investigate the nature of this QP Ising transition, as we expect it to be very different from the transitions described in this letter --- in particular, it likely has a finite dynamical exponent $z$, as a consequence of the prefactor $\kappa=1$ in the RG rules.  

\paragraph{\bf Discussion.}

We analyzed the critical behavior of quantum phase transitions separating ferromagnetic and paramagnetic phases in the quasiperiodic $q$-state Potts model in two dimensions. Using a controlled real-space renormalization group approach, we found that the critical behavior is independent of $q$, and is controlled by a new RG fixed point providing the first example of ``infinite-quasiperiodicity'' behavior in two dimensions. We argued on general grounds that such QP quantum phase transitions have correlation length exponent $\nu=1$, saturating a modified version of the Harris-Luck criterion. It would be interesting to find other examples of infinite-quasiperiodicity transitions, both in two and three dimensions. The case of the 2d QP Ising model also deserves more attention, as it should provide a different type of QP transition with finite dynamical exponent.  We leave these questions for future works.

\paragraph{Acknowledgments:} The authors thank Snir Gazit, Sid Parameswaran and Jed Pixley for useful discussions. This work was supported by the National Science Foundation under NSF Grant No. DMR-1653271 (S.G.), the US Department of Energy, Office of Science, Basic Energy Sciences, under Early Career Award No. DE-SC0019168 (U.A. and R.V.), and the Alfred P. Sloan Foundation through a Sloan Research Fellowship (R.V.)

\bibliography{2dQP,diss_ising}

\bigskip
\onecolumngrid
\newpage


\section*{Supplementary material}

# Supplemental material for "Quantum criticality in the 2d quasiperiodic Potts model"

Utkarsh Agrawal and Romain Vasseur

*Department of Physics, University of Massachusetts, Amherst, MA 01003, USA*

Sarang Gopalakrishnan

*Department of Physics and Astronomy, CUNY College of Staten Island,*
*Staten Island, NY 10314; Physics Program and Initiative for the Theoretical Sciences,*
*The Graduate Center, CUNY, New York, NY 10016, USA*

## I. CRITICAL BEHAVIOR FOR DIFFERENT VALUES OF $q$ AND QUASIPERIODIC POTENTIALS

In the main text we focused on the critical properties for a given quasiperiodic (QP) distribution (eq 2 in the main text) for $q = 3$. Here we present some numerical results for different distributions and values of $q$, illustrating the universality of our results.

### A. q > 3 Potts model

Fig. 1 shows the collapse of the magnetization for $q = 6$ and $q = 10$, for the QP potentials studied in the main text. The exponents quoted in the main text also lead to very good collapses in these cases, strongly suggesting a single universality class for all values of $q$. We also find that the gap scales as $-\ln \Delta E \sim \ln^2 L$ (not shown), as for $q = 3$.

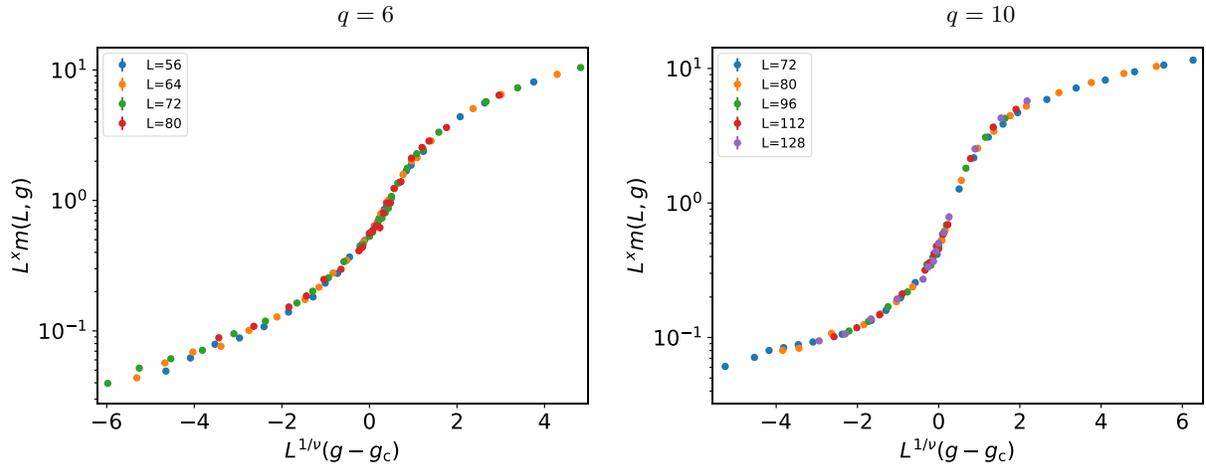

FIG. 1: Magnetization collapses for $q = 6$ and $q = 10$. The critical exponents $\nu = 1$ and $x = 0.92$ used in these plots are the same as for $q = 3$.

### B. Different angle in the QP potentials

In the main text we focused on the angle $\theta = \sqrt{2}\pi$ in the quasiperiodic potentials. Fig. 2 shows the magnetization collapse for $\theta = \sqrt{3}\pi$, again with the same critical exponents.

### C. Different QP potentials

In the main text, we defined the QP potentials using the logarithmic variables $-\ln J_{ij}$ and $-\ln h_i$ to decrease the transient nature of the RG. We show here that a more natural choice for the the QP potentials expressed directly in terms of $J_{ij}, h_i$ leads to the same critical behavior, albeit with stronger finite size effects. This is expected from our

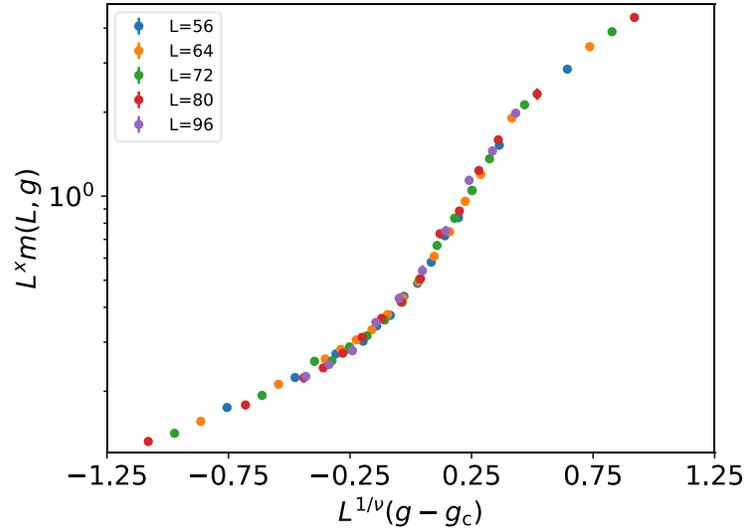

FIG. 2: Collapse for $\theta = \sqrt{3}\pi$ for $q = 3$ with the critical exponents $\nu = 1, x = 0.92$

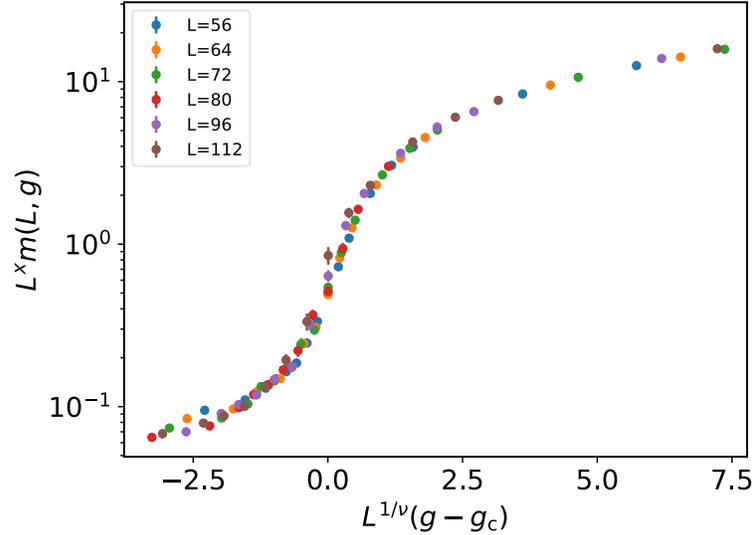

FIG. 3: Magnetization collapse for QP potentials defined in (1) with $g := -\ln \lambda$. The critical exponents are $\nu = 1$ and $x = 0.92$ as in the main text.

argument for $\nu = 1$, that relies on the repetition of defects which should scale universally for most functions with frequency $\varphi$. We consider

$$
\begin{aligned}
J_{ij} &= 4 + \cos\left(2\pi\varphi\vec{k}_1.\vec{r} + \phi_1\right) + \cos\left(2\pi\varphi\vec{k}_2.\vec{r} + \phi_2\right) \\
h_i &= \lambda(4 + \cos\left(2\pi\varphi i_x + \phi_3\right) + \cos\left(2\pi\varphi i_y + \phi_4\right)),
\end{aligned}
\tag{1}
$$

with the various variables defined in the main text below eq 2. The parameter tuning the transition is defined as $g \equiv -\ln \lambda$. Fig. 3 shows a good magnetization collapse using the same critical exponents as in the main text.

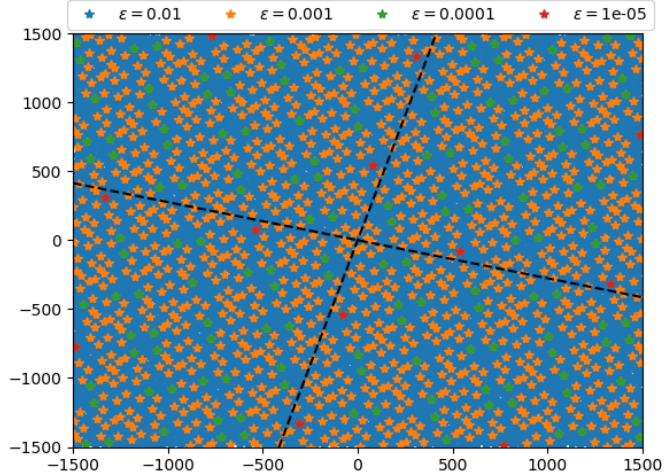

FIG. 4: The set of points $(x, y)$ such that $0 < f(x, y) < \epsilon$ (with $f(x, y) = 2 + \cos\left(2\pi\varphi\vec{k}_1.\vec{r}\right) + \cos\left(2\pi\varphi\vec{k}_2.\vec{r}\right)$, see text) for various values of $\epsilon$. The dashed lines represent the orientation of $\vec{k}_1$ and $\vec{k}_2$. As $\epsilon$ is decreased, the typical length scale separating points in this set increases.

## II. REPETITION PATTERNS IN 2D QUASIPERIODIC FUNCTIONS

To complement the discussion in the main text regarding the repetition of "defects", we consider a simplified question that illustrates this pattern. Given a 2d QP function $f(x, y) \geq 0$, we consider the pattern of the points $(x, y)$ such that $0 < f(x, y) < \epsilon$, as a function of $\epsilon$. We choose $f(x, y) = 2 + \cos\left(2\pi\varphi\vec{k}_1.\vec{r}\right) + \cos\left(2\pi\varphi\vec{k}_2.\vec{r}\right)$ and $k_1 = (\sin\theta, \cos\theta)$, with $\theta = \sqrt{5}\pi/2$; and plot the set $M_\epsilon = \{(x, y) \mid 0 < f(x, y) < \epsilon\}$ in Fig. 4 for various values of $\epsilon$. We see that the set $M_\epsilon$ has some clear structure, with the typical length scale separating points in the set $M_\epsilon$ increasing with decreasing $\epsilon$. We find similar behavior for different angles $\theta$, phases, and frequencies; however, the pattern of the set $M_\epsilon$ is not universal.

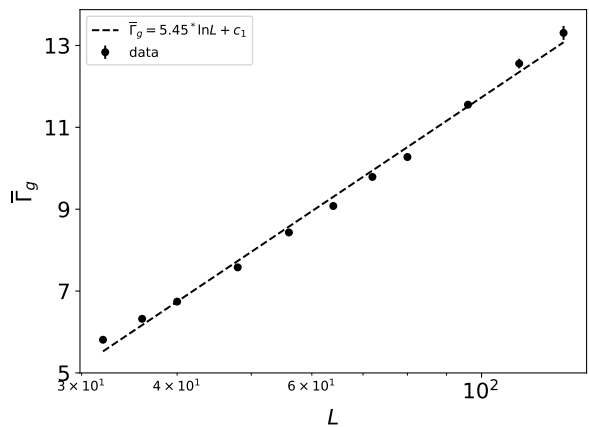

FIG. 5: Linear fit of the (logarithm of the) gap $\Gamma_g = -\ln \Delta E$ vs $\ln L$, possibly consistent with a large but finite dynamical exponent $z \approx 5.45$. Note the upward curvature trend indicating a faster than any power law decay of $\Delta E$.

### III. SCALING OF THE GAP

#### A. Additional numerical data

In the main text we argued that the logarithmic of the gap, $-\ln \Delta E \equiv \Gamma_g$, scales with system size as $\Gamma_g \sim \ln^2 L$. Here we try to fit the finite-size gap $\Delta E$ as a power law of system size, $\Delta E \sim L^{-z}$ (i.e finite dynamical exponent). The results are shown in Fig. 5, with the best fit giving $\Delta E \sim L^{-5.45}$, corresponding to a fairly large dynamical exponent $z \approx 5.45$. Although we cannot definitely rule out such a finite, large dynamical exponent, we do observe a slight upward curvature in Fig. 5, consistent with a logarithmically-diverging apparent dynamical exponent $z(L) \sim \ln L \to \infty$.

#### B. Bound on the scaling of the gap

In fact, we can rule out a finite dynamical exponents based on the general structure of the RG rules, assuming some self-similar structure at criticality. The banded structure of the couplings suggests that there is some fraction of initial couplings and fields (defined as $-\ln J$ and $-\ln h$ in the main text) which are local maxima and can be decimated in one go, in any order. Let us refer to the new set of bonds after decimating the local maxima as the first generation. After the above set of decimations we can again identify local maxima and decimate them to get 2nd generation couplings, fields and so on. Each generation eliminates certain fraction of spins. We assume that at the critical point this fraction is a constant in the sense that it is of same order for all generations. This implies that decimating a region of linear size $L$ to a single spin will require in general $\sim \log L$ generations. This structure can be made rigorous in one spatial dimension, but the banded structure of the couplings in our numerics strongly suggests that the same holds true in 2D as well.

We study the scaling of the gap with $L$ or equivalently with the number of generations $m$. After $m$ generations the typical magnetic fields $h^{(m)}$ will be given by $h^{(m)} = \frac{\prod^{M+1} h^0}{\kappa^{n_m} \prod^M J^0}$, where $h^{(0)}, J^{(0)}$ are values of fields and couplings in the initial distribution or the 0th generation. We show that the term $\kappa^{n_m}$ in the above expression gives a bound on the gap scaling and is enough to give an "infinite" dynamical exponent. Since we are only interested in the factors of $\kappa$, let us introduce the notation $\#[f]$ as the power of the $\kappa$ in the expression for $f$, where $f$ is a product/ratio of fields and couplings. We then prove that at the critical point we have $\#[h^{(m)}] \gtrsim m^2$ or in other words after running the RG upto length scale $L$, the typical magnetic field would decrease at least by a factor of $\sim \kappa^{-\log^2 L}$.

Moving on to the proof, we work with logarithmic variables $\ell_{[J,h]} \equiv -\ln[J, h]$. We have $\ell_h^{(m+1)} = \ell_h^{(m)} + \sum^n (\ell_h^{(m)} - \ell_J^{(m)}) + n \ln \kappa$, or $\#[\ell_h^{(m+1)}] = n + \#[\ell_h^{(m)}] + \#[\sum^n (\ell_h^{(m)} - \ell_J^{(m)})]$, where $n$ is some constant not depending on $m$. Note that each $\ell_h^{(m)}$ in the sum is greater than $\ell_J^{(m)}$ by construction. For each term in the sum we have $\#[\ell_h^{(m)} - \ell_J^{(m)}] = \#[\sum(\ell_h^{(m-1)} - \ell_J^{(m-1)})] + n_1$, where $n_1 > 0$ is because of the fact that typically to form a field greater than a coupling we need more decimations (at the critical point there is an asymmetry between $h$ and $J$). The value of $n_1$ is not important other than the fact that at the critical point we expect it to not depend on $m$, though it might show some fluctuations. This immediately gives $\#[h^{(m)} - \ell_J^{(m)}] \geq n_1 m$. Putting this in the expression for $\#[\ell_h^{(m+1)}]$ we get $\#[\ell_h^{(m+1)}] \geq n + \#[\ell_h^{(m)}] + n n_1 m$, or in other words, the above recursion implies $\#[\ell_h^{(m)}] \gtrsim m^2$.

This argument indicates that the gap at the critical point of the RSRG studied in this paper is bounded by $\kappa^{-\log^2 L}$, which comes from keeping track of the factors of $\kappa$ along the RG. For the Ising model ($\kappa = 1$) we do not see signs of "infinite" dynamical constant or of infinite quasiperiodic behavior, suggesting that the above lower bound is saturated. This suggest the saturation of the $\log^2 L$ scaling for the Potts model ($\kappa > 1$) as shown in a figure in the main text.



\end{document}